\documentclass[runningheads]{llncs}
\usepackage{amsmath}
\usepackage{amssymb}
\usepackage{amsfonts}
\usepackage{graphicx}
\usepackage{float}
\usepackage{subfigure}
\usepackage{booktabs}
\usepackage{multirow}
\usepackage{diagbox}
\usepackage{stfloats}
\usepackage{cite}
\usepackage{url}
\usepackage{xspace}
\usepackage{siunitx}
\usepackage{todonotes}
\usepackage[misc]{ifsym}

\usepackage[marginal]{footmisc}

\makeatletter

\newcommand{\Rmnum}[1]{\expandafter\@slowromancap\romannumeral #1@}
\makeatother

\begin{document}
\title{Blind Image Quality Assessment for MRI with A Deep Three-dimensional content-adaptive Hyper-Network}
\titlerunning{HyS-net}

\author{
    Kehan Qi \inst{*1,2} 
    \and Haoran Li \inst{*1,2} 
    \and Chuyu Rong\inst{1,2}
    \and Yu Gong\inst{1,4} 
    \and Cheng Li\inst{1}
    \and Hairong Zheng\inst{1}
    \and Shanshan Wang\inst{1,3 (\textrm{\Letter})}
}

\authorrunning{K. Qi et al.}
\institute{Paul C. Lauterbur Research Center for Biomedical Imaging, Shenzhen Institutes of Advanced Technology, Chinese Academy of Sciences, Shenzhen, Guangdong, China \email{sophiasswang@hotmail.com} 
\and University of Chinese Academy of Sciences, Beijing, China
\and Pengcheng Laboratory, Shenzhen, Guangdong, China
\and College of Medicine and Biological Information Engineering, Northeastern University, Shenyang, Liaoning, China}

\maketitle

\footnote{* These authors contribute equally to this work.\\}

\begin{abstract}
Image Quality Assessment (IQA) is of great value in the workflow of Magnetic Resonance Imaging (MRI)-based analysis. Blind IQA (BIQA) methods are especially required since high-quality reference MRI images are usually not available. Recently, many efforts have been devoted to developing deep learning-based BIQA approaches. However, the performance of these methods is limited due to the utilization of simple content-non-adaptive network parameters and the waste of the important 3D spatial information of the medical images. To address these issues, we design a 3D content-adaptive hyper-network for MRI BIQA. The overall 3D configuration enables the exploration of comprehensive 3D spatial information from MRI images, while the developed content-adaptive hyper-network contributes to the self-adaptive capacity of network parameters and thus, facilitates better BIQA performance. The effectiveness of the proposed method is extensively evaluated on the open dataset, MRIQC. Promising performance is achieved compared with the corresponding baseline and 4 state-of-the-art BIQA methods. We make our code available at \url{https://git.openi.org.cn/SIAT_Wangshanshan/HyS-Net}.
\end{abstract}

\section{Introduction}
Magnetic Resonance Imaging (MRI) plays a crucial role in many clinical applications such as medical diagnosis \cite{bruno2019new}, disease staging \cite{taylor2019diagnostic} and clinical research \cite{debette2019clinical} since it can produce highly detailed images with excellent soft-tissue contrast. It is worth noting that the quality of MRI images has a decisive influence on the diagnosis of the radiologist in the clinic. Low-quality MRI images are inevitable since MRI image quality is affected by various factors, for example, artifact and noise. MRI Image Quality Assessment (IQA) can improve the accuracy of clinical diagnosis. However, high-quality reference MRI images are usually not available in clinical practices for the development of full-reference IQA method. Therefore, there is an urgent demand of MRI non-reference IQA or blind IQA (BIQA) models to help the radiologist make more accurate diagnoses.

Currently, BIQA can be divided into two categories: methods for specific types of distortions and methods for general types of distortions. Methods belonging to the first category are designed to detect certain types of distortions, such as ringing effects \cite{liu2009no}, artifacts \cite{li2013referenceless}, blur \cite{li2015no} and so on. The prerequisite of these methods is that only known types of distortions exist in the image. However, it may not in line with the reality condition. Therefore, more and more researchers tend to develop methods for general types of distortions. Generally, this category of methods transforms the BIQA problem into a classification or regression problem, where classification and regression solvers are trained through extracting image features \cite{ghadiyaram2017perceptual, gu2014using, mittal2012no, saad2012blind, xu2016blind, xue2014blind, xue2013learning, ye2012unsupervised, zhang2015feature}. There are two kinds of methods for image feature extraction. One explores image features by calculating the natural scene statistics of the image \cite{saad2012blind}, and the other accomplishes the task through learning-based approaches \cite{zhang2015feature}. An overall assumption of the former is that the certain statistical laws shown in the image can be used for distortion evaluation. In other words, these methods estimate the image quality by extracting features that can indicate the degree of deviations of statistics in the distorted image. An example method of this kind is BLINDS-II \cite{ma2016group}. One major disadvantage of these methods is the slow processing speed as the hand-craft image conversion procedure is computationally time-consuming. Recently, deep learning-based methods have found wide employment in many computer vision tasks \cite{wang2019dimension, chen2019model, wang2020deepcomplexmri, yang2019clci, qi2019x, zhou2018radiomics} including BIQA \cite{bosse2017deep, kang2014convolutional, lin2018hallucinated, ma2017end, yan2019naturalness, zhang2018blind, zhang2020learning, su2020blindly}. Deep learning models, especially convolutional neural networks (CNN), can directly utilize the original images as model inputs to extract image features as well as perform the quality assessment end-to-end. CNNs can effectively learn complex mappings with minimal domain knowledge. However, the performance of current deep learning-based BIQA methods is limited. On the one hand, the parameters of the network are content-free, which forms a gap between model predictions and human perception. For example, human inspectors will consider an image of a clear blue sky of high quality while the most of BIQA methods will consider it of low quality since there is a large flattened area. Therefore, predicting image quality before image content understanding does not conform to the rule of humans assessing the quality of the image. On the other hand, most existing methods are 2D. The application of 2D approaches on 3D images wastes important spatial image information and deteriorates the performance of IQA.

To address these two issues mentioned above, we propose a 3D spatial-related hyper-network-based MRI BIQA method dubbed as HyS-net. The 3D spatial-related structure is designed to comprehensively explore 3D spatial information of the 3D images. Furthermore, we develop a content-adaptive hyper-network to introduce dynamics to the network parameters and help achieve better BIQA performance. The contributions of this paper are as follows: 
\begin{enumerate}
    \item [1)] A Hyper-network-based 3D Spatial-related model (HyS-net) is designed for MRI BIQA. It introduces 3D spatial information from MRI images for performance improvement.
    \item [2)] A self-adaptive parameter generating capacity is enabled by a developed content-adaptive hyper-network. It bridges the gap between the BIQA method and human perception.
    \item [3)] The proposed method has been extensively evaluated on an open dataset. Compared to state-of-the-art methods, the proposed HyS-net achieves promising performance.
\end{enumerate}

\section{Method}
\begin{figure*}[!htbp]
\centering
\includegraphics[width=1\linewidth]{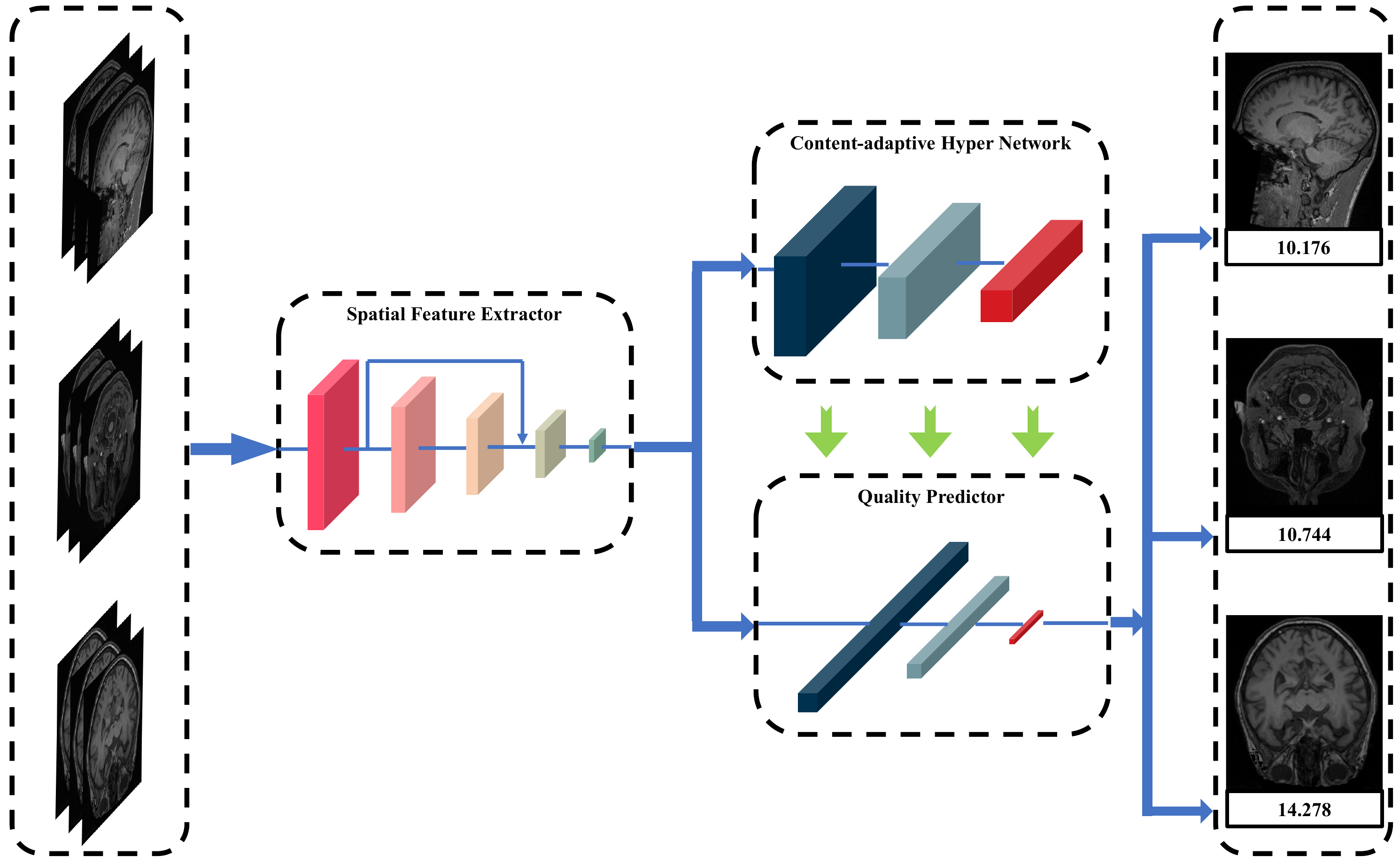}
\caption{The architecture of the proposed HyS-net. There are three parts: spatial feature extractor, content-adaptive hyper-network, and quality predictor. The blue line and green arrow indicates the data flow and parameter assignment operator respectively.}
\label{Fig.1}
\end{figure*}

In this paper, we develop a BIQA method that incorporates spatial information exploration and image content understanding to achieve better quality assessment performance. As shown in Fig. \ref{Fig.1}, the proposed HyS-net consists of three parts. The first part is a backbone network that serves as a spatial feature extractor. The second part is a content-adaptive hyper-network that is designed to generate dynamic parameters for the following part. The parameters are generated by understanding the content of the input image. The third part is a fully connected network-based quality predictor. In the following, we will introduce the details of the proposed HyS-net and the corresponding sub-networks.

\subsection{Spatial Feature Extractor}
Commonly, the image feature extractor explores shallow image features from a single image, such as texture, contrast, shape, and so on. Although this kind of feature exploration gains better performance than handcrafted approaches, it can still be futher improved for MRI BIQA. Complementary information exists in 3D MRI images, for example, structure and component. Therefore, it is necessary to introduce the 3D spatial information for MRI BIQA. 3D feature extractors such as 3D Res-net \cite{hara2018can} are introduced in our approach. The function of the spatial feature extractor can be described as:
\begin{align}
    \label{Eq.4}
        s=\mathcal{S}(\widehat{x}, \eta)
\end{align}
where $\mathcal{S}$, $\widehat{x}$, and $\eta$ denote the mapping function, multi adjacent slices input, and corresponding parameters of the spatial feature extractors, respectively. As shown in Fig. \ref{Fig.1}, the spatial feature extractor locates at the head of the proposed HyS-net.

\subsection{Self-adaptive BIQA Model}
The procedure of traditional deep learning-based BIQA methods can be described as:
\begin{align}
    \label{Eq.1}
        q=\psi(x, \theta)
\end{align}
where $q$, $\psi$, $x$, and $\theta$ denotes the image quality score, the quality prediction model, input image, and model weight parameters, respectively. Note that the model weight parameters are fixed after the training process. In other words, the network extracts the same kind of image features from different images. As illustrated in \cite{li2018has}, humans perform the image quality assessment after understanding the content of the images. For better BIQA performance, the rule of image quality prediction of the model should also adjust according to the varying image content. To this end, we model the task of BIQA as: 
\begin{align}
    \label{Eq.2}
        \psi(x, \theta_{x})=q
\end{align}
The difference between $\theta$ and $\theta_{x}$ is that $\theta_{x}$ varies for different input images. $\theta_{x}$ serves as a self-adaptive image quality prediction rule. Specifically, we design a content-adaptive hyper-network to achieve the purpose:
\begin{align}
    \label{Eq.3}
        \theta_{x}=\mathcal{H}(\mathcal{S}(x), \gamma)
\end{align}
where $\mathcal{H}$ and $\gamma$ denotes the mapping function and parameters of the content-adaptive hyper-network, respectively. $\mathcal{S}(x)$ denotes the input of the content-adaptive hyper-network, i.e. the image features extracted by the spatial feature extractor from image $x$. The mapping function $\mathcal{H}$ refers to the mapping from image content to image quality prediction. After introducing the intermediate variable $\theta_{x}$ and the hyper-network $\mathcal{H}$, we can divide the workflow of the proposed HyS-net into three-step: spatial feature extraction, self-adaptive image quality prediction rule learning, and image quality prediction. A top-down image quality assessment mechanism that conducts the image content understanding in the first place bridges the gap between the BIQA method and human perception. Besides, it improves the adaptability of the network when facing large image content differences.

\subsection{Content-adaptive Hyper-network}
\begin{figure*}[!htbp]
\centering
\includegraphics[width=1\linewidth]{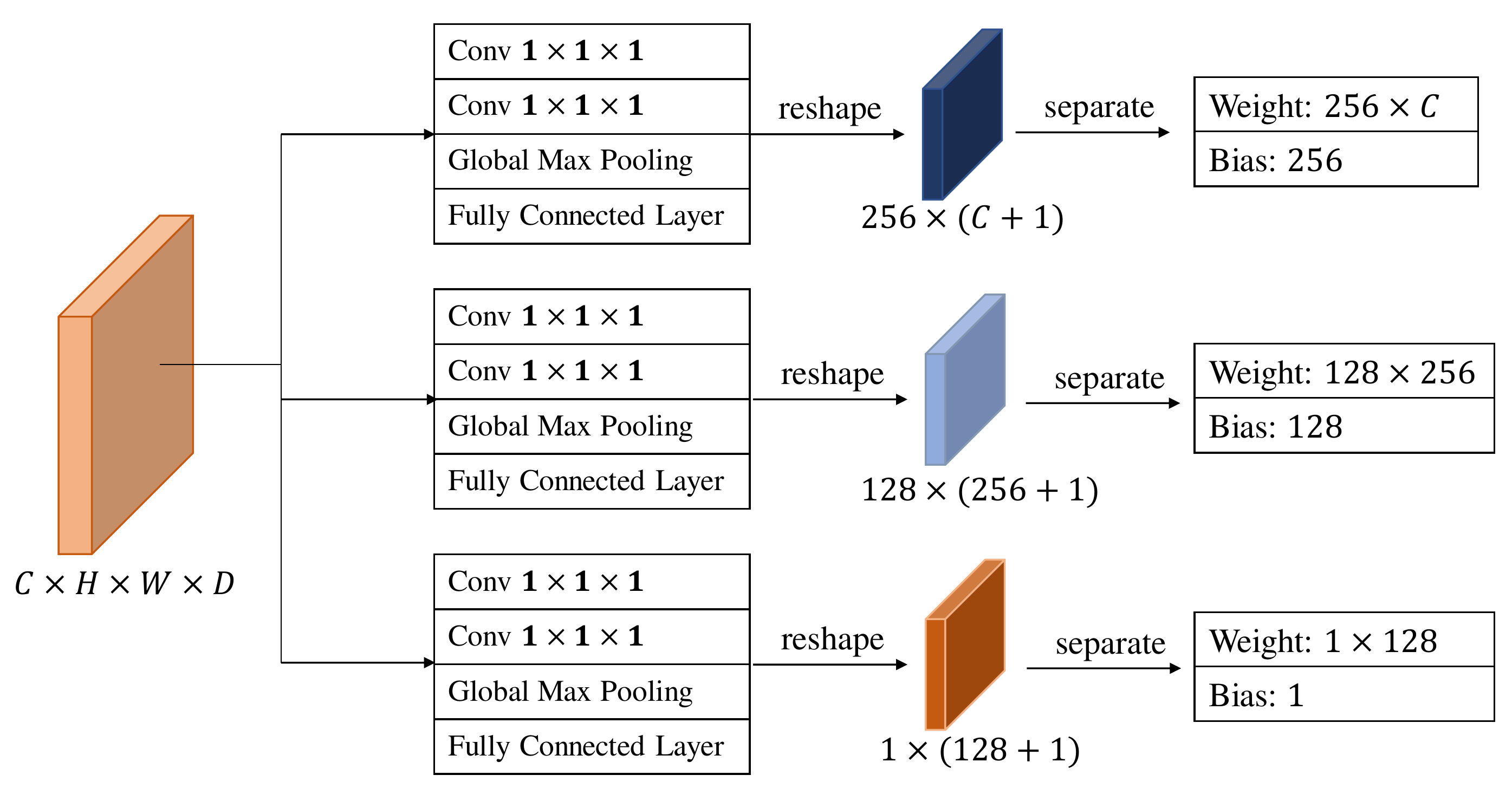}
\caption{The pipeline of our content-adaptive hyper-network. Given an input tensor with shape $C\times H\times W\times D$, where $C$ indicates the number of channels, our hyper-network produces weights and bias for the three target fully connected layers.}
\label{Fig.2}
\end{figure*}

As shown in Fig. \ref{Fig.2}, our content-adaptive hyper-network $\mathcal{H}(\cdot, \gamma)$ consists of 3 branches to generate the parameters for three fully connected layers. Each branch consists of two $1\times 1\times 1$ convolution layers, one Global Max Pooling layer, and one fully connected layer. The target parameters (including weights and bias) are generated from the last fully connected layer followed by a reshape operation. The generated weights and bias are then employed as parameters of corresponding fully connected network(FCN)-based quality predictor and instructs the subsequent image quality prediction.

\subsection{Quality Predictor}
After extracting the spatial features and generating the content-adaptive parameters, we use a simple fully-connected-layer-based quality predictor to map the spatial features to quality score. As shown in Fig. \ref{Fig.1}, this predictor consists of three fully-connected layers, which is fed by the features extracted by the spatial feature extractor, and employ weights and bias generated by hyper-network.  

\subsection{Implementation Detail}
All the 3D MRI images are normalized to $(0,1]$. For spatial feature extraction, we randomly sample 9 patches with a size of $96\times 96\times 96$ during training and orderly sample all the possible patches with a step size of 48 and a patch size of $96\times 96\times 96$ during testing. For non-spatial feature extraction in ablation analysis, we resize each slice to $384\times 384$ by padding the borders. 
The model was implemented using Pytorch deep learning framework with NVIDIA 2080Ti GPUs. We used $L_1$ loss function for training. Adam optimizer was adopted to train our model for 50 epochs. Learning rate was firstly set to $1e-4$, and reduced by by a factor of 2 after every 10 epochs. 

\section{Experimental Results}
\subsubsection{Dataset}
Magnetic Resonance Imaging Quality Control (MRIQC) is a MRI image quality assessment research project initiated by Stanford University \cite{esteban2019crowdsourced}. It aims at producing and sharing resource of image quality metrics and annotations from researchers to train human experts and automated algorithms. Currently, MRIQC provides more than 50,000 T1-weighted and T2-weighted MRI images. In this study, we used 3,257 downloadable T1-weighted images with different types of quality scores provided. We employ the \textit{total Signal-to-Noise Ratio (SNR)} as our measure of image quality. 1629 cases were randomly selected for training, with the rest 1628 cases reserved for testing.

\subsubsection{Quantitative Metric}
In this study, we calculate two quantitative metrics, Spearman’s Rank-Order Correlation Coefficient (SROCC) and Pearson’s Linear Correlation Coefficient (PLCC), to evaluate the performance of MRI BIQA. A value close to 1 for SROCC and PLCC indicates a better quality assessment result.

\subsubsection{Ablation Experiments}
Ablation experiments are designed to analyze the effects of the different components of HyS-Net, including the structure of the backbone network, the introduction of the content-adaptive hyper-network, and the use of the spatial feature. The details and the quantitative metrics of the experimental results are shown in Table \ref{Table.1}. 


Surprisingly, results indicate that simply increasing the network depth (from Res-net 18 to Res-net 50) reduces the network performance. Introducing the spatial features and content-adaptive hyper-network can effectively enhance the network performance. 
The content-adaptive hyper-network and spatial features are complementary. Using the content-adaptive hyper-network and spatial features at the same time can maximize the performance improvement of the network. 
However, it is worth noting that the performance improvement is marginal and sometimes, the performance becomes even worse when introducing the content-adaptive hyper-network only. The reason could be that the contexts in the 2D domain cannot provide sufficient information for content-adaptive hyper-network. Nevertheless, when spatial information is extracted with the introduction of the 3D spatial feature extractor, the performance can be effectively elevated.

\begin{table*}[!htbp]
	\caption{The details and the quantitative metrics of image quality assessment results of the models involved in the ablation experiments. For each metric, we mark the best in bold.}
	    \label{Table.1}
	        \centering
                \begin{tabular}{ccccc}
                    \toprule
Baseline & Hyper-network & 3D Spatial & SROCC & PLCC \\
\multicolumn{1}{c}{\multirow{4}{*}{Res-net 18}} &               &                 & 0.7324 & 0.7996 \\
\multicolumn{1}{c}{}                            & $\checkmark$  &                 & 0.7196 & 0.8034 \\
\multicolumn{1}{c}{}                            &               &   $\checkmark$  & 0.7374 & 0.8096 \\
\multicolumn{1}{c}{}                            & $\checkmark$  &   $\checkmark$  & \textbf{0.7567} & \textbf{0.8226} \\
\hline
\multirow{4}{*}{Res-net 34}                     &               &                 & 0.7106 & 0.7592 \\
                                                & $\checkmark$  &                 & 0.7298 & 0.7997 \\
                                                &               &   $\checkmark$  & 0.7409 & 0.8128 \\
                                                & $\checkmark$  &   $\checkmark$  & \textbf{0.7519} & \textbf{0.8208} \\
\hline                                            
\multirow{4}{*}{Res-net 50}                     &               &                 & 0.6863 & 0.7697 \\
                                                & $\checkmark$  &                 & 0.6867 & 0.7171 \\
                                                &               &   $\checkmark$  & 0.7448 & 0.8169 \\
                                                & $\checkmark$  &   $\checkmark$  & \textbf{0.7566} & \textbf{0.8231} \\
                    \bottomrule
                \end{tabular}
\end{table*}

\subsubsection{Comparison to State-of-the-art Methods}
To further validate the performance of the proposed HyS-net, we also reimplemented 4 recently published state-of-the-art BIQA methods including BRISQUE \cite{mittal2012no}, DBCNN \cite{zhang2018blind}, deepIQA \cite{bosse2017deep}, and Two Stream CNN \cite{yan2018two}. 
Results are listed in Table \ref{Table.2}. It can be summarized that our proposed model performs better than all the four methods. SROCC and PLCC are respectively increased by 0.0307 and 0.0234 with our method. These results confirm the effectiveness of our proposed method and the promising applications of our method in clinical applications. 

\begin{table*}[!htbp]
	\caption{The quantitative metrics of image quality assessment results of the models involved in the comparison experiments. For each metric, we mark the best in bold.}
	    \label{Table.2}
	        \centering
                \begin{tabular}{ccccc}
                    \toprule
               &       SROCC     &       PLCC      \\
HyS-net(ours)        & \textbf{0.7566} & \textbf{0.8231} \\
BRISQUE \cite{mittal2012no} &      0.3483     &      0.5849     \\
DBCNN \cite{zhang2018blind} &      0.7121     &      0.7719     \\
deepIQA \cite{bosse2017deep} &      0.7259     &      0.7997     \\
Two Stream CNN \cite{yan2018two} &      0.7230     &      0.7898     \\
                    \bottomrule
\end{tabular}
\end{table*}

\subsubsection{Visualization of Generated Dynamic Weights}
\begin{figure*}[!htbp]
\centering
\includegraphics[width=1\linewidth]{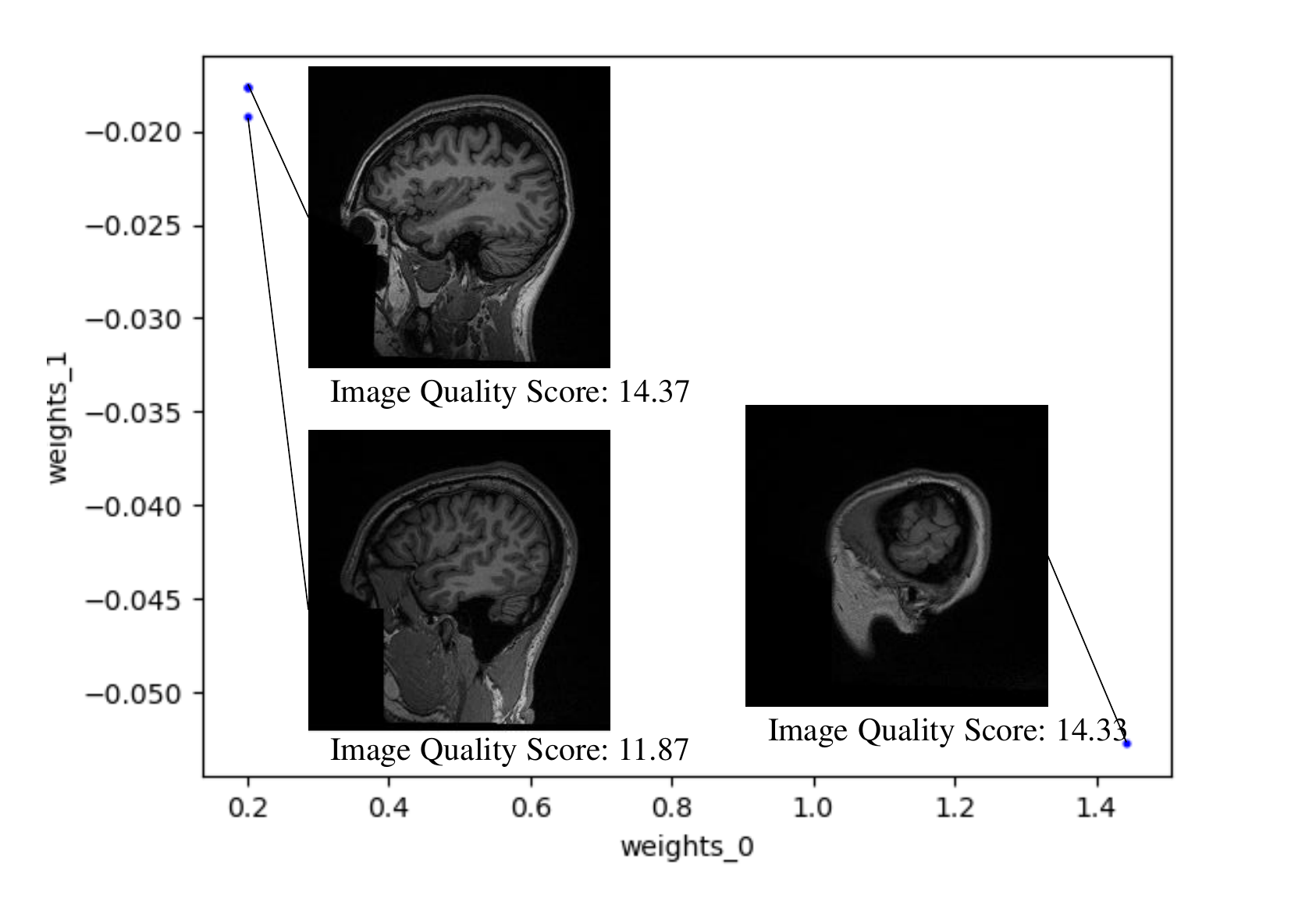}
\caption{Weights of regressor generated by our hyper-network from different images plotted in the 2D space after PCA. This figure illustrates that generated weights are content-adaptive, even if the image quality of input images are similar.}
\label{Fig.4}
\end{figure*}

To verify the effectiveness of the weight generating procedure, weights generated by the hyper-network are extracted. These weights come from images of varied contents. After that, PCA transformation are adopted to analyze the weights. Finally, they are plotted in the 2D space for visualization (Fig. \ref{Fig.4}). This figure indicates that the generated weights vary for different image contents, and the hyper-network generates similar weights for similar image objects. Thus, our hyper-network achieves a content-adaptive network weights for BIQA.

\section{Conclusion}
In this paper, we present a deep learning model, named HyS-net, for MRI Blind Image Quality Assessment (BIQA). HyS-net can explore the spatial information from 3D MR images through the overall 3D configuration. Furthermore, it enables self-adaptive capacity of network parameters by content-adaptive hyper-network, which contributes to better BIQA performance. The proposed method gracefully addresses the problems of existing approaches - the lack of spatial information and content-non-adaptive model parameters. Experiments on the MRIQC dataset demonstrates that our proposed HyS-net could achieve promising performance compared with existing methods.



\begin{thebibliography}{40}
\bibitem{bruno2019new}
Bruno, Federico, et al. ``New advances in {MRI} diagnosis of degenerative osteoarthropathy of the peripheral joints." La radiologia medica 124.11 (2019): 1121-1127.
\bibitem{taylor2019diagnostic}
Taylor, Stuart A., et al. ``Diagnostic accuracy of whole-body {MRI} versus standard imaging pathways for metastatic disease in newly diagnosed non-small-cell lung cancer: the prospective streamline {L} trial." The Lancet Respiratory Medicine 7.6 (2019): 523-532.
\bibitem{debette2019clinical}
Debette, Stéphanie, et al. ``Clinical significance of {M}agnetic {R}esonance {I}maging markers of vascular brain injury: a systematic review and meta-analysis." JAMA neurology 76.1 (2019): 81-94.
\bibitem{liu2009no}
Liu, Hantao, Nick Klomp, and Ingrid Heynderickx. ``A no-reference metric for perceived ringing artifacts in images." IEEE Transactions on Circuits and Systems for Video Technology 20.4 (2009): 529-539.
\bibitem{li2013referenceless}
Li, Leida, et al. ``Referenceless measure of blocking artifacts by {T}chebichef kernel analysis." IEEE Signal Processing Letters 21.1 (2013): 122-125.
\bibitem{li2015no}
Li, Leida, et al. ``No-reference image blur assessment based on discrete orthogonal moments." IEEE transactions on cybernetics 46.1 (2015): 39-50.
\bibitem{ghadiyaram2017perceptual}
Ghadiyaram, Deepti, and Alan C. Bovik. ``Perceptual quality prediction on authentically distorted images using a bag of features approach." Journal of vision 17.1 (2017): 32-32.
\bibitem{gu2014using}
Gu, Ke, et al. ``Using free energy principle for blind image quality assessment." IEEE Transactions on Multimedia 17.1 (2014): 50-63.
\bibitem{mittal2012no}
Mittal, Anish, Anush Krishna Moorthy, and Alan Conrad Bovik. ``No-reference image quality assessment in the spatial domain." IEEE Transactions on image processing 21.12 (2012): 4695-4708.
\bibitem{saad2012blind}
Saad, Michele A., Alan C. Bovik, and Christophe Charrier. ``Blind image quality assessment: {A} natural scene statistics approach in the {DCT} domain." IEEE transactions on Image Processing 21.8 (2012): 3339-3352.
\bibitem{xu2016blind}
Xu, Jingtao, et al. ``Blind image quality assessment based on high order statistics aggregation." IEEE Transactions on Image Processing 25.9 (2016): 4444-4457.
\bibitem{xue2014blind}
Xue, Wufeng, et al. ``Blind image quality assessment using joint statistics of gradient magnitude and {L}aplacian features." IEEE Transactions on Image Processing 23.11 (2014): 4850-4862.
\bibitem{xue2013learning}
Xue, Wufeng, Lei Zhang, and Xuanqin Mou. ``Learning without human scores for blind image quality assessment." Proceedings of the IEEE Conference on Computer Vision and Pattern Recognition. 2013.
\bibitem{ye2012unsupervised}
Ye, Peng, et al. ``Unsupervised feature learning framework for no-reference image quality assessment." 2012 IEEE conference on computer vision and pattern recognition. IEEE, 2012.
\bibitem{zhang2015feature}
Zhang, Lin, Lei Zhang, and Alan C. Bovik. ``A feature-enriched completely blind image quality evaluator." IEEE Transactions on Image Processing 24.8 (2015): 2579-2591.
\bibitem{ma2016group}
Ma, Kede, et al. ``Group mad competition-a new methodology to compare objective image quality models." Proceedings of the IEEE Conference on Computer Vision and Pattern Recognition. 2016.
\bibitem{wang2019dimension}
Wang, Shanshan, et al. ``{DIMENSION}: dynamic {MR} imaging with both k‐space and spatial prior knowledge obtained via multi‐supervised network training." NMR in Biomedicine (2019): e4131.
\bibitem{chen2019model}
Chen, Yanxia, et al. ``Model-based convolutional de-aliasing network learning for parallel {MR} imaging." International Conference on Medical Image Computing and Computer-Assisted Intervention. Springer, Cham, 2019.
\bibitem{wang2020deepcomplexmri}
Wang, Shanshan, et al. ``Deepcomplex{MRI}: {E}xploiting deep residual network for fast parallel {MR} imaging with complex convolution." Magnetic Resonance Imaging 68 (2020): 136-147.
\bibitem{yang2019clci}
Yang, Hao, et al. ``{CLCI}-Net: {C}ross-level fusion and context inference networks for lesion segmentation of chronic stroke." International Conference on Medical Image Computing and Computer-Assisted Intervention. Springer, Cham, 2019.
\bibitem{qi2019x}
Qi, Kehan, et al. ``X-net: {B}rain stroke lesion segmentation based on depthwise separable convolution and long-range dependencies." International Conference on Medical Image Computing and Computer-Assisted Intervention. Springer, Cham, 2019.
\bibitem{zhou2018radiomics}
Zhou, Yongjin, et al. ``A radiomics approach with {CNN} for shear-wave elastography breast tumor classification." IEEE Transactions on Biomedical Engineering 65.9 (2018): 1935-1942.
\bibitem{bosse2017deep}
Bosse, Sebastian, et al. ``Deep neural networks for no-reference and full-reference image quality assessment." IEEE Transactions on image processing 27.1 (2017): 206-219.
\bibitem{kang2014convolutional}
Kang, Le, et al. ``Convolutional neural networks for no-reference image quality assessment." Proceedings of the IEEE conference on computer vision and pattern recognition. 2014.
\bibitem{lin2018hallucinated}
Lin, Kwan-Yee, and Guanxiang Wang. ``Hallucinated-{IQA}: {N}o-reference image quality assessment via adversarial learning." Proceedings of the IEEE Conference on Computer Vision and Pattern Recognition. 2018.
\bibitem{ma2017end}
Ma, Kede, et al. ``End-to-end blind image quality assessment using deep neural networks." IEEE Transactions on Image Processing 27.3 (2017): 1202-1213.
\bibitem{yan2019naturalness}
Yan, Bo, Bahetiyaer Bare, and Weimin Tan. ``Naturalness-aware deep no-reference image quality assessment." IEEE Transactions on Multimedia 21.10 (2019): 2603-2615.
\bibitem{zhang2020learning}
Zhang, Weixia, et al. ``Learning to blindly assess image quality in the laboratory and wild." 2020 IEEE International Conference on Image Processing (ICIP). IEEE, 2020.
\bibitem{li2018has}
Li, Dingquan, et al. ``Which has better visual quality: {T}he clear blue sky or a blurry animal?." IEEE Transactions on Multimedia 21.5 (2018): 1221-1234.
\bibitem{hara2018can}
Hara, Kensho, Hirokatsu Kataoka, and Yutaka Satoh. ``Can spatiotemporal 3{D} {CNN}s retrace the history of 2{D} {CNN}s and imagenet?." Proceedings of the IEEE conference on Computer Vision and Pattern Recognition. 2018.
\bibitem{esteban2019crowdsourced}
Esteban, Oscar, et al. ``Crowdsourced {MRI} quality metrics and expert quality annotations for training of humans and machines." Scientific data 6.1 (2019): 1-7.
\bibitem{kim2016fully}
Kim, Jongyoo, and Sanghoon Lee. ``Fully deep blind image quality predictor." IEEE Journal of selected topics in signal processing 11.1 (2016): 206-220.
\bibitem{zhang2018blind}
Zhang, Weixia, et al. ``Blind image quality assessment using a deep bilinear convolutional neural network." IEEE Transactions on Circuits and Systems for Video Technology 30.1 (2018): 36-47.
\bibitem{yan2018two}
Yan, Qingsen, Dong Gong, and Yanning Zhang. ``Two-stream convolutional networks for blind image quality assessment." IEEE Transactions on Image Processing 28.5 (2018): 2200-2211.
\bibitem{su2020blindly}
Su, Shaolin, et al. ``Blindly assess image quality in the wild guided by a self-adaptive hyper network." Proceedings of the IEEE/CVF Conference on Computer Vision and Pattern Recognition. 2020.
\end{thebibliography}
\end{document}